\documentclass{article}
\usepackage{amssymb}

\usepackage{graphicx}
\usepackage{amsmath}

\input{tcilatex}

\begin{document}

\title{Quantum Decoherence Phenomena in the Thermal Assisted Tunneling}
\author{V. V. Makhro \\
Bratsk Industrial Institute, Bratsk, 665709, Russia\\
Phone: +7 3953 484724\\
E-mail: maxpo@excite.com}
\maketitle

We propose a novel approach to the problem of a transition from quantum to
classical behavior in mesoscopic spin systems. This paper is intended to
demonstrate that\ \ main cause of such transitions is quantum decoherence
which appear as a result of a thermal interaction between spin system and
its environment. We shall consider semiclassical model where spin problem
has been mapped onto particle problem. In such a case, particle will be
localized in one of two metastable potential wells, and transitions between
this states can occur either due to over-barrier transition or via quantum
tunneling. It is necessary to emphasize that thermal activation couldn't
appear in pure form as a classical phenomena even in semiclassical limit:
escape is determined not only by tunneling but also by the over-barrier
reflection in the case when particle obtain from the environment energy,
large then height of the barrier. We shell use the next computational
scheme. Consider an ensemble of particles localized in potential wells. Due
to interaction with environment, each of the particles can obtain some
energy $E$. It is clear that in this case , either due to tunneling or due
to over-barrier transition, probability of escape $P$ will be a function of
corresponding magnitude $E$: $P=P(E)$. If population for given $E$ in
ensemble is $n(E)$ than for the total escape probability one obtain  $P_{t}=%
\frac{1}{N}\int n(E)P(E)dE$, where $N$ is a number of particles in ensemble.
Since population $n(E)$ is also a function of a temperature $T$ we can
consider $P_{t}$ as a function of $T$ too. We want to emphasize again that
for such consideration, $P_{t}(T)$ will be absolutely smooth, without any
sharp turns, as long as $n(T)$ and $P(E)$ is smooth function.

But this picture strikingly changes when one take into account the
possibility of the coherence destruction by the interaction with
environment. It is well known that environment induces a dynamical
localization of the quantum state into a generalized coherent state. In
particular, in the double-well potential, for specific values of external
field's \ strength and frequency, tunneling is coherently destroyed, i.e., a
localized packet can be built as a superposition of two degenerate states,
which remains localized ''forever'' in one well. We study the behavior of
semiclassical system interacting with a classical environment represented by
an infinite set of harmonic oscillators with a frequencies $\varpi .$ For
each particle in our ensemble escape probability will be now\cite{espan} $%
P^{\prime }=PJ_{0}(\frac{2V}{\omega })$, where $J_{0}(x)$ is the zero-order
Bessel function. $V$, the magnitude of interaction with a bath, is defined
as $V=Q(\varpi )\varpi C,$ where $Q(\varpi )$ is a population for the
frequency $\varpi $, and $C$ is a coupling constant, which is an adjusting
parameter in considered case.

\FRAME{ftbpF}{5.1698in}{2.6558in}{0pt}{}{}{figmakhro.gif}{\special{language
"Scientific Word";type "GRAPHIC";maintain-aspect-ratio TRUE;display
"USEDEF";valid_file "F";width 5.1698in;height 2.6558in;depth
0pt;original-width 5.1145in;original-height 2.6143in;cropleft "0";croptop
"1";cropright "1";cropbottom "0";filename 'figMakhro.gif';file-properties
"XNPEU";}}

Using computational scheme described above, one can now to calculate total
escape probability with taking into account decoherence effects. In Fig. 1
we have plotted for example results of such a calculations for the
mesoscopic system $CrNi_{6}$ (curve 1). It is visible, that quantum
decoherence essentially changes behavior of $P_{t}(T)$ at low temperatures:
in contrast to usual thermal assisted tunneling (monotonic curve 2) the
decoherence leads to the origin of the irregularities of $P_{t}(T)$ for $%
T\lessapprox 10$ K. In Fig.1 we also plotted for comparison results of
experiments\cite{keren} (signed as ''+''), and results of computations for
the ''pure'' classical thermal activation (curve 3). Similarity between
curve 1 and experimental data seems to be very encouraging.


\begin{thebibliography}{9}
\bibitem{espan}  J.M. Gomes Llorente and J. Plata, Phys. Rev. A, \textbf{45}%
, 6960 (1992).

\bibitem{keren}  A.Keren, P. Mendels, A.Kratzer, A. Scuiller, M. Verdauger,
Z. Slaman and C.Baines, cond-mat/9806230v3 (1998)
\end{thebibliography}
\end{document}